\documentstyle[a4,12pt]{article}
\newcommand{\delfour}{{\Delta^{(4)}}}
\newcommand{\delsq}{\Delta^{(2)}}
\newcommand{\xv}{\vec{x}}
\newcommand{\yv}{\vec{y}}
\newcommand{\mbare}{M^{0}_{c}}
\newcommand{\sigmav}{\mbox{\boldmath$\sigma$}}
\newcommand{\NRQCDcoll}{
C.~T.~H.~Davies,$^a$
K.~Hornbostel,$^b$
G.~P.~Lepage,$^c$ \\
A.~J.~Lidsey,$^a$
J.~Shigemitsu,$^d$
J.~Sloan$^e$ \\[.4cm]

\small $^a$University of Glasgow, Glasgow, UK G12 8QQ. UKQCD collaboration.\\
\small $^b$Southern Methodist University, Dallas, TX 75275. \\
\small $^c$Newman Laboratory of Nuclear Studies, Cornell University,
Ithaca, NY 14853. \\
\small $^d$The Ohio State University, Columbus, OH 43210. \\
\small $^e$Florida State University, SCRI, Tallahassee, FL 32306.
}

\begin{document}

\title{Precision Charmonium Spectroscopy {}From Lattice QCD }

\author{
\NRQCDcoll \\ }

\maketitle

\begin{abstract}
\noindent
We present results for Charmonium spectroscopy using
Non-Relativistic QCD (NRQCD). For the NRQCD action the
leading order spin-dependent and next to leading
order spin-independent interactions have been included with
tadpole-improved coefficients.
We use multi-exponential fits to multiple correlation
functions to extract ground and excited $S$ states.
Splittings between the lowest $S$, $P$ and $D$ states
are given and
we have accurate values for the $S$ state hyperfine splitting
and the $\chi_c$ fine structure.
Agreement with experiment is good - the
remaining systematic errors are discussed.
\end{abstract}

\section{Introduction}
The study of heavy-heavy mesons is important for Lattice Gauge
Theory not only because of the availability of experimental
data for comparison but also because such systems allow a quantitative
study of systematic errors which arise in lattice simulations at present.
To study heavy-heavy mesons we use Non-Relativistic QCD
 (NRQCD)\cite{nrqcd,cornell} and
previously we have reported a very successful study of the Bottomonium
system\cite{spectrum}.
 This allowed the extraction of two fundamental
parameters in QCD\cite{poster}, the b-quark mass \cite{bmass}
and the strong coupling constant
$\alpha_{s}$ \cite{alphas}.
 Here we report on a similar study of the Charmonium
spectrum.

The starting point of NRQCD is to expand the original QCD
lagrangian in powers of $v^{2}$, the typical quark velocity in
a bound state.
For the $J/\Psi$ system $v^{2} \sim 0.3$. Thus we
systematically include relativistic errors order by order in
$v^{2}$ away
from a Non-Relativistic limit. Our action is
the same one as used in \cite{spectrum} where relativistic corrections
${\cal O} (Mv^{4})$ have been included. This means that systematic errors
from relativistic corrections
will be ${\cal O} (M_{c}v^{6})$ (= $\approx$ $30 - 40$~ MeV) for the $J/\Psi$
system i.e. 10\% in spin-independent
 splittings and 30\% in
spin-dependent splittings. This is considerably less
accurate than for the $\Upsilon$ case\cite{spectrum} because
$v^{2}$ is about a factor of 3 larger here.
Other sources of systematic error include
discretisation errors and errors from
the absence of virtual quark loops
because we use
quenched configurations generated with the standard
plaquette action. Finite volume errors
should be negligible because of the relatively small size of the $J/\Psi$
system.

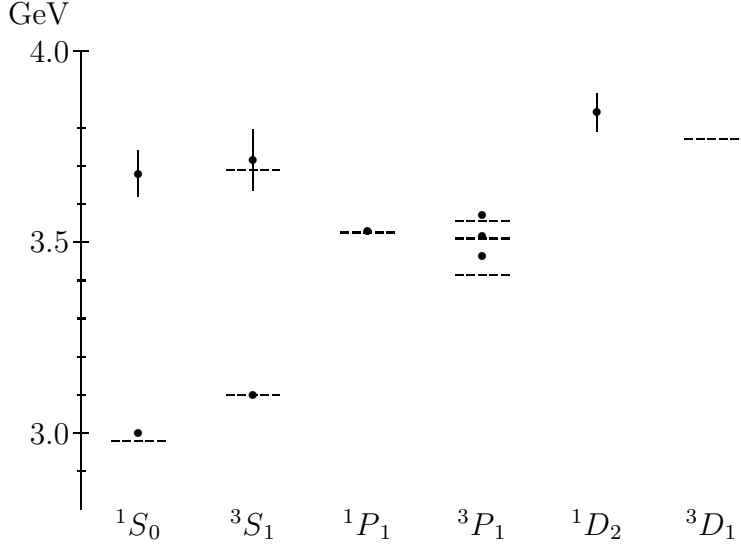
\begin{figure}[t]
\begin{center}
\setlength{\unitlength}{.02in}
\begin{picture}(200,140)(0,280)
\put(15,280){\line(0,1){120}}
\multiput(13,300)(0,50){3}{\line(1,0){4}}
\multiput(14,290)(0,10){10}{\line(1,0){2}}
\put(12,300){\makebox(0,0)[r]{3.0}}
\put(12,350){\makebox(0,0)[r]{3.5}}
\put(12,400){\makebox(0,0)[r]{4.0}}
\put(12,410){\makebox(0,0)[r]{GeV}}

     \put(30,280){\makebox(0,0)[t]{${^1S}_0$}}
\multiput(23,298)(3,0){5}{\line(1,0){2}}
     \put(30,300){\circle*{2}}

     \put(30,368){\circle*{2}}
     \put(30,368){\line(0,1){6}}
     \put(30,368){\line(0,-1){6}}

     \put(60,280){\makebox(0,0)[t]{${^3S}_1$}}
\multiput(53,310)(3,0){5}{\line(1,0){2}}
     \put(60,310){\circle*{2}}

\multiput(53,369)(3,0){5}{\line(1,0){2}}
     \put(60,371.5){\circle*{2}}
     \put(60,371.5){\line(0,1){8}}
     \put(60,371.5){\line(0,-1){8}}

     \put(90,280){\makebox(0,0)[t]{${^1P}_1$}}
\multiput(83,352.6)(3,0){5}{\line(1,0){2}}
     \put(90,353){\circle*{2}}
     \put(90,353){\line(1,0){1}}
     \put(90,353){\line(0,-1){1}}

     \put(120,280){\makebox(0,0)[t]{${^3P}_1$}}
\multiput(113,355.6)(3,0){5}{\line(1,0){2}}
     \put(120,357.2){\circle*{2}}
\multiput(113,351)(3,0){5}{\line(1,0){2}}
     \put(120,351.8){\circle*{2}}
\multiput(113,341.5)(3,0){5}{\line(1,0){2}}
     \put(120,346.4){\circle*{2}}

     \put(150,280){\makebox(0,0)[t]{${^1D}_2$}}
     \put(150,384){\circle*{2}}
     \put(150,384){\line(0,1){5}}
     \put(150,384){\line(0,-1){5}}

     \put(180,280){\makebox(0,0)[t]{${^3D}_1$}}
\multiput(173,377)(3,0){5}{\line(1,0){2}}


\end{picture}
\end{center}
 \caption{NRQCD simulation results for the spectrum of the
$J/\Psi$ system
using an
inverse lattice spacing of 1.23~GeV, fixed from the
spin-averaged 1P-1S splitting. The $^1S_0$ mass is fixed
at 3.0 GeV, from a fit to the kinetic mass.
  Experimental values are indicated by dashed lines.
 Error bars are shown where visible, and only indicate statistical
 uncertainties.}
\label{fig:cspinav}
\end{figure}

\begin{figure}[tb]
\begin{center}
\setlength{\unitlength}{0.013in}
\begin{picture}(200,200)(0,-140)
\put(15,-130){\line(0,1){180}}
\multiput(13,-120)(0,40){5}{\line(1,0){4}}
\multiput(14,-120)(0,10){16}{\line(1,0){2}}
\put(12,-120){\makebox(0,0)[r]{$-120$}}
\put(12,-80){\makebox(0,0)[r]{$-80$}}
\put(12,-40){\makebox(0,0)[r]{$-40$}}
\put(12,0){\makebox(0,0)[r]{$0$}}
\put(12,40){\makebox(0,0)[r]{$40$}}
\put(12,50){\makebox(0,0)[r]{MeV}}


\multiput(28,30)(3,0){5}{\line(1,0){2}}
     \put(57,30){\makebox(0,0)[t]{$J/\Psi$}}
     \put(35,24){\circle*{3}}

\multiput(28,-89)(3,0){5}{\line(1,0){2}}
     \put(54,-84){\makebox(0,0)[t]{$\eta_c$}}
     \put(35,-72){\circle*{3}}

\multiput(83,1)(3,0){5}{\line(1,0){2}}
     \put(98,1){\makebox(0,0)[l]{$h_c$}}
     \put(90,-9){\circle*{3}}
     \put(90,-9){\line(0,1){2}}
     \put(90,-9){\line(0,-1){2}}

\multiput(123,-110)(3,0){5}{\line(1,0){2}}
     \put(139,-110){\makebox(0,0)[l]{$\chi_{c0}$}}
     \put(130,-78){\circle*{3}}
     \put(130,-78){\line(0,1){6}}
     \put(130,-78){\line(0,-1){6}}

\multiput(123,-15)(3,0){5}{\line(1,0){2}}
     \put(139,-15){\makebox(0,0)[l]{$\chi_{c1}$}}
     \put(130,-24){\circle*{3}}
     \put(130,-24){\line(0,1){4}}
     \put(130,-24){\line(0,-1){4}}

\multiput(123,31)(3,0){5}{\line(1,0){2}}
     \put(139,31){\makebox(0,0)[l]{$\chi_{c2}$}}
     \put(130,30){\circle*{3}}
     \put(130,30){\line(0,1){4}}
     \put(130,30){\line(0,-1){4}}
\end{picture}
\end{center}
 \caption{Simulation results for the spin structure of the
 $J/\Psi$~family, using an inverse lattice
 spacing of 1.23~GeV.
The energies of the spin-averaged S and P states have been
set to zero.
  Error bars for points are statistical.}

\label{fig:cspinp}
\end{figure}
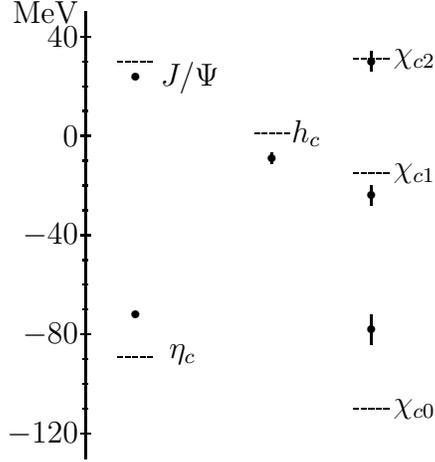

Shown in Figures (\ref{fig:cspinav}) and (\ref{fig:cspinp})
is the spectrum for Charmonium using Lattice NRQCD.
The spectrum was calculated using an ensemble of 273 gauge
field configurations generated with the standard Wilson
action at $\beta$ = 5.7\cite{thanks_ukqcd}.
To set the scale we fix our simulation
result for the spin-averaged 1P-1S splitting to its experimental value of
458 MeV.
This gives $a^{-1}$ = 1.23(4) GeV, where the uncertainty
is purely statistical. Since we are working in the
quenched approximation this value can be and is different
both from that obtained at the same value of $\beta$
using light hadron spectroscopy\cite{weingarten}
or using Upsilon spectroscopy\cite{us_in_progress}.
We expect a value fixed from heavyonium to be more
accurate than that from light hadron spectroscopy
because spin-independent splittings in the heavy quark
sector are independent of quark mass to a good approximation and
systematic errors are under better control\cite{nrqcd}.

To fix the bare quark mass in the action, $M^{0}_{c}$,
we plot a dispersion relation correct up to ${\cal O} (v^{4})$
for the $\eta_c$. $M^{0}_{c}$ is then tuned
until the simulation value for the
kinetic mass is equal to the experimental value
of the mass of the $\eta_c$ (2.98 GeV). We find that using $aM^{0}_{c}$=0.8
gives $M(\eta_c)$=3.0(1) GeV with $a^{-1}$ = 1.23(4) GeV.

In Figure (\ref{fig:cspinav}) the whole
Charmonium spectrum is shown and in Figure (\ref{fig:cspinp}) the
spin-dependent splittings are shown in more detail. In Figure
(\ref{fig:cspinp}) it can be seen that although the general pattern of
splittings for the S and P states is reproduced well, systematic errors
are visible above the statistical errors. It should then be
possible in the future to observe systematic improvements to the
current calculation, when higher order relativistic corrections are
included and further discretisation and quenching errors are
removed.

We give details in section 2 of our evolution equation
and the quark Greens function used to make up meson correlation functions.
Section 3 describes the results from the simulation
using multi-exponential fits.
We  illustrate the need for
multiple smearing functions
to obtain smaller statistical errors. Section 4 compares simulation
results to experiment and section 5 contains our conclusion.

\section{Evolution Equation and Quark propagators}

One of the advantages of the formulation of NRQCD is that it
involves a simple difference equation in the temporal direction.
This allows the evolution of the quark Green function
as an initial value
problem which can be solved with one sweep through the lattice.
We define our quark Green function to be initially
\begin{eqnarray}
 G_1 &=&
  \left(1\!-\!\frac{aH_0}{2n}\right)^{n}
 U^\dagger_4
 \left(1\!-\!\frac{aH_0}{2n}\right)^{n} \, \delta_{\xv,0}
\end{eqnarray}
and then continue to evolve using
\begin{eqnarray}
  G_{t+1} &=&
  \left(1\!-\!\frac{aH_0}{2n}\right)^{n}
 U^\dagger_4
 \left(1\!-\!\frac{aH_0}{2n}\right)^{n}\left(1\!-\!a\delta H\right) G_t
 \quad (t>1) .
\label{tevolve}
\end{eqnarray}
On the lattice, the kinetic energy operator is
 \begin{equation}
 H_0 = - {\delsq\over2\mbare},
 \end{equation}
and the correction terms are
 \begin{eqnarray}
\label{deltaH}
\delta H
&=& - c_1 \frac{(\delsq)^2}{8(\mbare)^3}
            + c_2 \frac{ig}{8(\mbare)^2}\left({\bf \Delta}\cdot{\bf E} -
{\bf E}\cdot{\bf \Delta}\right) \nonumber \\
 & & - c_3 \frac{g}{8(\mbare)^2} \sigmav\cdot({\bf \Delta}\times{\bf E} -
{\bf E}\times{\bf \Delta})
 - c_4 \frac{g}{2\mbare}\,\sigmav\cdot{\bf B}  \nonumber \\
 & &  + c_5 \frac{a^2\delfour}{24\mbare}
     -  c_6 \frac{a(\delsq)^2}{16n(\mbare)^2} .
\end{eqnarray}
The first two terms in $\delta H$ are spin-independent
relativistic corrections and the next two are spin-dependent
correction terms which contribute to
 the P and S spin splittings respectively. The last two terms
come from finite lattice spacing
corrections to the lattice Laplacian and the lattice time derivative.
${\bf \Delta}$ is the symmetric lattice derivative, $\delsq$
is the lattice form of the Laplacian and
 $\delfour$ is a lattice version of the continuum
operator $\sum D_i^4$. We used the standard traceless
cloverleaf operators for the
chromo-electric and magnetic fields, $\bf E$ and~$\bf B$.  The
parameter~$n$
is introduced to remove instabilities in the heavy quark
propagator caused by the highest momentum modes of the theory\cite{nrqcd}.
For our simulations at $\beta = 5.7$ and with a bare mass for the
$c$ quark in lattice units of 0.8, we set $n = 4$.

The coupling constants $c_{i}$ appearing in
equation (\ref{deltaH}) can be calculated by matching NRQCD to
full QCD \cite{cornell,colin}. At tree level all the coefficients are one.
The largest radiative corrections are believed to be
tadpole contributions\cite{maclep}.
We take care of these by using the
method suggested in \cite{maclep} where all the U's are redefined by
\begin{eqnarray}
\label{US}
U_{\mu}(x) \rightarrow \frac{U_{\mu}(x)}{u_{0}}
\end{eqnarray}
with $u_{0}$ the fourth root of the plaquette (at $\beta$=5.7 we
use $u_0$ = 0.861). Since the cloverleaf
expression involves the evaluation of a plaquette this renormalization
will have the effect of redefining
${\bf E}$ and ${\bf B}$ via
\begin{eqnarray}
\label{EB}
{\bf E} \rightarrow \frac{{\bf E}}{u^{4}_{0}} \hspace{15pt}
{\bf B} \rightarrow \frac{{\bf B}}{u^{4}_{0}}
\end{eqnarray}
which will strongly affect spin-dependent splittings.
With the dominant tadpole contributions thus removed, we
use the tree level values for the $c_{i}$'s. The only remaining free
parameters are the bare quark mass $M^{0}_{c}$ and
the bare coupling constant $g$ which appear
in the original QCD Lagrangian. All the details of the quark evolution
up to this point are identical to those in \cite{spectrum}. In the
following some of the technical details differ slightly.

Given the quark propagators in equation (\ref{tevolve})
it is relatively straightforward
to combine them appropriately to form meson propagators
with specific quantum numbers. Using the notation of \cite{spectrum}
we take $\psi^{\dag}$ to create a heavy quark and $\chi^{\dag}$ to
create a heavy anti-quark. Then the following interpolating
operator creates a meson centred on the point $\vec{x}_1$ :
\begin{equation}
 \sum_{\xv_2} {\psi^{\dagger}}(\xv_1) \, \Gamma(\xv_1 - \xv_2) \, \chi^\dagger
(\xv_2).
\label{meson}
\end{equation}
Local meson operators are tabulated in \cite{spectrum}. Here we
generalise the operators to include `smearing functions'. For
S states the meson operator $\Gamma$ becomes $\Omega \; \phi(\vec{x_1} -
\vec{x_2})$ where
$\Omega$ is a $2 \times 2$ matrix in spin space giving the
quantum numbers of the meson and $\phi$ is a
simple approximation to the wavefunction.
For P states, $\phi$ also becomes a $p$ wavefunction, which can
be thought of as the derivative of a spherically symmetric
function \cite{spectrum}. In general $\Gamma$ is a sum of
spin matrices multiplying different smearing functions, generalising
the operators in \cite{spectrum}.
For the wavefunctions $\phi$ we use here
 wavefunctions from a $1/r$ potential with
their spread adjusted to match the size of the appropriate meson.

For meson propagators at zero momentum we then have
\begin{equation}
 G_{meson}(\vec{p}=0,t) =
 \sum_{\yv_1,\yv_2}Tr \left[ G^\dagger_t(\yv_2) \,
{\Gamma^{(sk)}}^\dagger (\yv_1 - \yv_2) \,
\tilde{G}_t(\yv_1) \right]
\label{gmes2}
\end{equation}
with
\begin{equation}
\tilde{G}_t(\yv) \equiv \sum_{\xv}G_t(\yv - \xv) \, \Gamma^{(sc)}(\xv).
\end{equation}
$\Gamma^{sc}(x)$ and $\Gamma^{sk}(x)$ refers to the meson operator
$\Gamma(x) = \Omega \phi(x)$ with the smearing function $\phi(x)$
at the source or sink respectively and enumerated by the
integer
$n_{sc}$ or $n_{sk}$. $n$ = 1 corresponds to the ground state
meson, $n$ = 2 to the first radially excited state. $\tilde{G}_t$ is obtained
using equations (1)
and (2) with $\delta_{\vec{x},0} \rightarrow \Gamma^{(sc)}(\vec{x})$.
The trace is over color
and spin. The convolutions are evaluated
using Fast Fourier Transforms.

We also study finite momentum propagators for the $^{1}S_0$ meson, given by:
\begin{equation}
 G_{meson}(\vec{p},t) =
 \sum_{\yv_1}Tr \left[ G^\dagger_t(\yv_1)\Omega
\tilde{G}_t(\yv_1) \right] e^{-i{\vec{p}} \cdot \yv_1}
\label{gmes}
\end{equation}

Using the notation $^{2S+1}L_{J}$, we have looked at meson
propagators for the following states: $^{1}S_{0}$, $^{3}S_{1}$, $^{1}P_{1}$,
$^{3}P_{0}$, $^{3}P_{1}$, $^{3}P_{2}$ for both the
E and T representation and the $^{1}D_{2}$ in the T representation.
For the
S states, smearing functions both for the ground and first radially excited
state were used as well as a local $\delta$ function ($n$ = $loc$). {}From
this all possible combinations of smearing at the source and sink
were formed
making a $3 \times 3$ matrix of
correlation functions. For the P and D states only the ground state
smearing function was used at the source.
We calculated the dispersion relation for the $^{1}S_{0}$
by looking at the meson propagator
for small momentum components using $(n_{sc},n_{sk}) = (loc,loc)$ and
$(1,loc)$.
To maximize our statistics we use all
color and spin indices at the source
when calculating our meson propagators.
For the $^{3}S_{1}$, $^{1}P_{1}$, $^{3}P_{1}$, $^{3}P_{2}$ and $^{1}D_{2}$ we
average over polarization directions making a total of 30 S, P and
D meson
propagators to analyze.

\section{\bf Simulation results}
In the simulation we used 273 quenched gluon field configurations
on a
$12^{3}$x24 lattice at $\beta = 5.7$
generously supplied by the UKQCD collaboration\cite{thanks_ukqcd}. They were
fixed
to Coulomb gauge using a Fourier accelerated steepest descents algorithm
\cite{facc} with a cutoff on $[\partial \cdot A]^2$ of $10^{-6}$.
Due to  the relatively small size of the $J/\Psi$ it is possible to use
more than one starting site on a spatial slice.
We also use more than one starting point in time to increase
statistics.
In this case we used 8 different
spatial origins and 2 different starting times at timeslice 1 and 12.
If we bin the spatial origins together we find significant correlation,
whereas binning together two propagators with an initial timeslice of
1 and 12 but with the same spatial origin
gives little or no correlation at all.
For most of our fits we bin together all the correlation functions
from a given configuration,
except when doing multiple-exponential multiple-correlation
fits for the $^{1}S_{0}$ and $^{3}S_{1}$ case. Here we only
 bin on spatial origin
and having the increased sample size from the time direction
significantly improves the fit. We also
checked, however, that fitting with all data unbinned produces
 a worse $\chi^{2}$ than
when all data is binned, another indicator of spatial correlations.

In NRQCD, as in QCD, there are two free parameters,
the bare coupling constant $g$ and the
bare quark mass $\mbare$. We fix $g$ implicitly when we set the
scale $a^{-1}$. To fix $\mbare$ we tune
so that the simulation result for the kinetic mass
of the $^{1}S_{0}$ agrees with the
experimental value of the mass of the $\eta_{c}$ (2.98 GeV).
For this we find $E_{\bf P}$ for several different
momenta of the $^{1}S_{0}$ and fit to the form
\begin{equation} \label{dispers}
E_{\bf P} - E_{0} = \frac{{\bf P}^{2}}{2M_{kin}} -
C_{1}\frac{({\bf P}^{2})^{2}}{8M^{3}_{kin}}
- \frac{C_{2}}{8M^{3}_{kin}}\sum_{i} {\bf P}^{4}_{i}
\end{equation}
simultaneously for momenta components (1,0,0), (1,1,0), (1,1,1),
(2,0,0) in units of $2\pi/12a$. $M_{kin}$ is taken to be the rest mass
of the $\eta_{c}$.
$C_1$ should take the value 1.0 in a fully Lorentz invariant theory.
Instead we find the value 1.7(1) - this is because of relativistic corrections
that have not been included. The mass in the ${\bf P}^4$ term differs
from that in the ${\bf P}^2$ term by the cube root of 1.7 i.e. 20\%.
Since we have included no relativistic corrections to the ${\bf P}^4$ term
we would expect it to be correct only to leading order i.e. 30\%, and
the difference we observe is consistent with that.
We expect results closer to 1.0 for the $\Upsilon$ because this is
a more non-relativistic system. Indeed there \cite{spectrum} we
find a tendency for $C_1$ to be larger than 1.0 but consistent with
1.0 within rather larger errors of size 0.3.
The last term in equation (\ref{dispers}) is
a non-rotationally-invariant term allowed on
the lattice but $C_2$
is found to be $-0.1(1)$
consistent with zero. This indicates that no
discretisation errors are visible in the dispersion relation
once the ${\cal O} (a^{2})$ terms in
the heavy quark action have been taken care of in equation
(\ref{deltaH}).
A fit with the extra ${\bf P}^{6}$ relativistic correction in
was tried and no significant signal for it was found.
Conversely a fit with
just the leading order
${\bf P}^{2}$ term in was tried but gave a very poor $Q$ value.
This suggests that with the particular
momentum components used a fit including terms up to ${\bf P}^{4}$
is appropriate.
Using a bare quark mass of $a\mbare$ = 0.8 gives
$M_{kin}a = 2.429(7)$ or
$M_{kin} = 3.0(1)$ GeV using $a^{-1}$ = 1.23(4) GeV.
 All simulation results quoted here are
from using this value of the bare quark mass.
The error on the bare quark mass is then of order 10\% from
both statistical errors in $a^{-1}$ and systematic errors
from higher order relativistic corrections.

In Figures (\ref{fig:splots}) and (\ref{fig:pplots}) we
show effective masses for the $^{1}S_{0}$ and $^{1}P_{1}$
states respectively.
We use the na\"{\i}ve definition
$m_{eff}(t) = -log(G(t+1)/G(t))$ together with bootstrap
errors. {}From the
S state plots it is clear that smearing has the effect
of producing an earlier plateau in the effective mass. Although the statistical
errors have increased for the smeared cases as compared to the
local-local case the earlier plateau allows fitting to take
place closer to the origin and ultimately produces better errors.
For the first excited state a plateau
cannot be seen for the effective mass
and the signal ultimately decays to the ground state.
A better transient plateau was seen for the excited S state
in the $\Upsilon$ spectrum at $\beta$ = 6.0~\cite{spectrum}. This reflects
the fact that at higher
$\beta$ values the excited states have smaller masses in lattice units and
last for longer times.
For the P state the signal/noise ratio is much poorer than that for the
S state, as expected.

\renewcommand{\Diamond}{}
\setlength{\unitlength}{0.240900pt}
\begin{figure}
\label{fig:splots}
\noindent

}
\put(950,-60){t}
}
\end{picture}

\caption{$^{1}S_{0}$ Effective masses by (source, sink).}

\end{figure}
\subsection{Fitting Results for the $^{1}S_{0}$ and $^{3}S_{1}$
and the singlet $P$ and $D$ states}
We use a variety of fitting routines to
extract high precision ground state
masses for the $^{1}S_{0}$ and $^{3}S_{1}$ as well as masses for their
first radially excited states. We have used in general the same
fitting procedures which are described in more detail in \cite{spectrum}.

Multi-exponential fits allow a fit to the correlation function
at much earlier times than single-exponential fits,
thus reducing the noise. A fit
to $n$ exponentials allows confidence in the masses of the
first $n-1$ states. Since, as described above, excited states die
very rapidly at low $\beta$, it is much harder to get a value for
an excited state mass at $\beta$=5.7 than at $\beta$ = 6.0. This is
reflected in our errors. It is also true, however, that the ground
state plateau appears earlier and the use of many exponentials
to get to early times is not as important at $\beta$ = 5.7 as at
$\beta$ = 6.0.

The first type of fit we do is that to a matrix of correlation functions:
\begin{equation}
G_{meson}(n_{sc},n_{sk};t) = \sum_{k=1}^{N_{exp}}\; a(n_{sc},k)\,a^*(n_{sk},k)
e^{-E_{k}\cdot  t}
\label{matansatz}
\end{equation}
For the S states we use the combination $n_{sc} = 1,2$ and
$n_{sk} = 1,2$ forming a 2 x 2 matrix. Then we perform fits for
$N_{exp}$ = 2 and 3. Our fitting procedure inverts the covariance matrix
using the svd algorithm. We have sufficiently good statistics that we are
able to keep all eigenvectors of the covariance matrix and achieve a
good fit\cite{bat-cd}.

For the second fit a row of correlation functions is formed
and fitted to
\begin{equation}
G_{meson}(n_{sc},loc;t) = \sum_{k=1}^{N_{exp}}\; b(n_{sc},k)\,
e^{-E_k \cdot t }
\label{mcorfit}
\end{equation}
We use the correlation functions $(n_{sc},n_{sk})$
with  $n_{sc} = 1,2$ $ n_{sk} = loc$.
Again fits use $N_{exp}$ = 2, 3.

In Tables \ref{1s0} and
\ref{3s1} are results from the row and matrix fits for the
$^{1}S_{0}$ and $^{3}S_{1}$.
The errors stated are those causing a change $\delta \chi ^{2} = 1$
and we also quote the quality of the fit, Q. For an acceptable fit
Q should be in the range 0.01 to 0.9  and ideally $Q > 0.1$.
To improve our statistics we only bin correlation functions which start
from different spatial origins but not ones which have different starting
timeslices.
This has little effect on the central value but does increase the Q
value giving us more confidence in the fit.

{}From both
tables it is clear that an accurate ground state mass can be obtained at
very early times. Only a $t_{min}$ of 2 gives an unacceptable Q for
the 2 exponential fit. Adding a 3rd exponential produces an
acceptable fit, although we don't take this value
because Q increases further as $t_{min}$ is increased.
 This contrasts with the higher $t_{min}$ needed
for $\Upsilon$ spectroscopy
at $\beta$ = 6.0\cite{spectrum}.
The masses we obtain
are independent of the type of fitting routine within errors, although
the values for Q are lower for the matrix fits.
At this point it is constructive to test how effective the multiple
exponential fits are for the ground states at $\beta$ = 5.7.
In Table \ref{1s0single} are values for a single exponential fit
to the $(n_{sc},n_{sk})$ = $(1,loc)$ and $(1,1)$ for the
$^{1}S_{0}$ state. In both cases an acceptable $Q$
requires $t_{min}$ of 6, significantly larger than for the multiple exponential
fit. We choose fitted values 0.6182(7) and 0.697(1) for the
$^1S_0$ and $^3S_1$ ground states respectively.

For
the first excited state the choice of fitted value is far more difficult.
To have confidence in the value we should use a 3 exponential fit
although this gives larger errors in the fitted masses.
We look for both a steady value in the fitted mass as $t_{min}$
 is changed and a
steady value for Q. It is also useful to look at the
amplitude for the second excited state in the 3 exponential
fit to see at what $t_{min}$ values it has decayed away.

\renewcommand{\Diamond}{}
\setlength{\unitlength}{0.240900pt}
\begin{figure}
\noindent
\begin{picture}(1500,500)
{
\put(200,0) {

\setlength{\unitlength}{0.240900pt}
\ifx\plotpoint\undefined\newsavebox{\plotpoint}\fi
\begin{picture}(750,540)(0,0)
\font\gnuplot=cmr10 at 10pt
\gnuplot
\sbox{\plotpoint}{\rule[-0.200pt]{0.400pt}{0.400pt}}%
\put(176.0,68.0){\rule[-0.200pt]{0.400pt}{108.164pt}}
\put(176.0,68.0){\rule[-0.200pt]{4.818pt}{0.400pt}}
\put(154,68){\makebox(0,0)[r]{0.8}}
\put(666.0,68.0){\rule[-0.200pt]{4.818pt}{0.400pt}}
\put(176.0,143.0){\rule[-0.200pt]{4.818pt}{0.400pt}}
\put(154,143){\makebox(0,0)[r]{0.9}}
\put(666.0,143.0){\rule[-0.200pt]{4.818pt}{0.400pt}}
\put(176.0,218.0){\rule[-0.200pt]{4.818pt}{0.400pt}}
\put(154,218){\makebox(0,0)[r]{1}}
\put(666.0,218.0){\rule[-0.200pt]{4.818pt}{0.400pt}}
\put(176.0,293.0){\rule[-0.200pt]{4.818pt}{0.400pt}}
\put(154,293){\makebox(0,0)[r]{1.1}}
\put(666.0,293.0){\rule[-0.200pt]{4.818pt}{0.400pt}}
\put(176.0,367.0){\rule[-0.200pt]{4.818pt}{0.400pt}}
\put(154,367){\makebox(0,0)[r]{1.2}}
\put(666.0,367.0){\rule[-0.200pt]{4.818pt}{0.400pt}}
\put(176.0,442.0){\rule[-0.200pt]{4.818pt}{0.400pt}}
\put(154,442){\makebox(0,0)[r]{1.3}}
\put(666.0,442.0){\rule[-0.200pt]{4.818pt}{0.400pt}}
\put(176.0,517.0){\rule[-0.200pt]{4.818pt}{0.400pt}}
\put(154,517){\makebox(0,0)[r]{1.4}}
\put(666.0,517.0){\rule[-0.200pt]{4.818pt}{0.400pt}}
\put(176.0,68.0){\rule[-0.200pt]{0.400pt}{4.818pt}}
\put(176,23){\makebox(0,0){0}}
\put(176.0,497.0){\rule[-0.200pt]{0.400pt}{4.818pt}}
\put(249.0,68.0){\rule[-0.200pt]{0.400pt}{4.818pt}}
\put(249,23){\makebox(0,0){2}}
\put(249.0,497.0){\rule[-0.200pt]{0.400pt}{4.818pt}}
\put(322.0,68.0){\rule[-0.200pt]{0.400pt}{4.818pt}}
\put(322,23){\makebox(0,0){4}}
\put(322.0,497.0){\rule[-0.200pt]{0.400pt}{4.818pt}}
\put(395.0,68.0){\rule[-0.200pt]{0.400pt}{4.818pt}}
\put(395,23){\makebox(0,0){6}}
\put(395.0,497.0){\rule[-0.200pt]{0.400pt}{4.818pt}}
\put(467.0,68.0){\rule[-0.200pt]{0.400pt}{4.818pt}}
\put(467,23){\makebox(0,0){8}}
\put(467.0,497.0){\rule[-0.200pt]{0.400pt}{4.818pt}}
\put(540.0,68.0){\rule[-0.200pt]{0.400pt}{4.818pt}}
\put(540,23){\makebox(0,0){10}}
\put(540.0,497.0){\rule[-0.200pt]{0.400pt}{4.818pt}}
\put(613.0,68.0){\rule[-0.200pt]{0.400pt}{4.818pt}}
\put(613,23){\makebox(0,0){12}}
\put(613.0,497.0){\rule[-0.200pt]{0.400pt}{4.818pt}}
\put(686.0,68.0){\rule[-0.200pt]{0.400pt}{4.818pt}}
\put(686,23){\makebox(0,0){14}}
\put(686.0,497.0){\rule[-0.200pt]{0.400pt}{4.818pt}}
\put(176.0,68.0){\rule[-0.200pt]{122.859pt}{0.400pt}}
\put(686.0,68.0){\rule[-0.200pt]{0.400pt}{108.164pt}}
\put(176.0,517.0){\rule[-0.200pt]{122.859pt}{0.400pt}}
\put(176.0,68.0){\rule[-0.200pt]{0.400pt}{108.164pt}}
\put(556,452){\makebox(0,0)[r]{(1,$\ell$)}}
\put(600,452){\raisebox{-.8pt}{\makebox(0,0){$\Diamond$}}}
\put(212,381){\raisebox{-.8pt}{\makebox(0,0){$\Diamond$}}}
\put(249,321){\raisebox{-.8pt}{\makebox(0,0){$\Diamond$}}}
\put(285,289){\raisebox{-.8pt}{\makebox(0,0){$\Diamond$}}}
\put(322,277){\raisebox{-.8pt}{\makebox(0,0){$\Diamond$}}}
\put(358,267){\raisebox{-.8pt}{\makebox(0,0){$\Diamond$}}}
\put(395,252){\raisebox{-.8pt}{\makebox(0,0){$\Diamond$}}}
\put(431,271){\raisebox{-.8pt}{\makebox(0,0){$\Diamond$}}}
\put(467,270){\raisebox{-.8pt}{\makebox(0,0){$\Diamond$}}}
\put(504,270){\raisebox{-.8pt}{\makebox(0,0){$\Diamond$}}}
\put(540,206){\raisebox{-.8pt}{\makebox(0,0){$\Diamond$}}}
\put(577,219){\raisebox{-.8pt}{\makebox(0,0){$\Diamond$}}}
\put(613,294){\raisebox{-.8pt}{\makebox(0,0){$\Diamond$}}}
\put(650,239){\raisebox{-.8pt}{\makebox(0,0){$\Diamond$}}}
\put(212.0,379.0){\rule[-0.200pt]{0.400pt}{0.964pt}}
\put(202.0,379.0){\rule[-0.200pt]{4.818pt}{0.400pt}}
\put(202.0,383.0){\rule[-0.200pt]{4.818pt}{0.400pt}}
\put(249.0,319.0){\rule[-0.200pt]{0.400pt}{1.204pt}}
\put(239.0,319.0){\rule[-0.200pt]{4.818pt}{0.400pt}}
\put(239.0,324.0){\rule[-0.200pt]{4.818pt}{0.400pt}}
\put(285.0,287.0){\rule[-0.200pt]{0.400pt}{1.204pt}}
\put(275.0,287.0){\rule[-0.200pt]{4.818pt}{0.400pt}}
\put(275.0,292.0){\rule[-0.200pt]{4.818pt}{0.400pt}}
\put(322.0,272.0){\rule[-0.200pt]{0.400pt}{2.168pt}}
\put(312.0,272.0){\rule[-0.200pt]{4.818pt}{0.400pt}}
\put(312.0,281.0){\rule[-0.200pt]{4.818pt}{0.400pt}}
\put(358.0,260.0){\rule[-0.200pt]{0.400pt}{3.132pt}}
\put(348.0,260.0){\rule[-0.200pt]{4.818pt}{0.400pt}}
\put(348.0,273.0){\rule[-0.200pt]{4.818pt}{0.400pt}}
\put(395.0,245.0){\rule[-0.200pt]{0.400pt}{3.373pt}}
\put(385.0,245.0){\rule[-0.200pt]{4.818pt}{0.400pt}}
\put(385.0,259.0){\rule[-0.200pt]{4.818pt}{0.400pt}}
\put(431.0,263.0){\rule[-0.200pt]{0.400pt}{3.854pt}}
\put(421.0,263.0){\rule[-0.200pt]{4.818pt}{0.400pt}}
\put(421.0,279.0){\rule[-0.200pt]{4.818pt}{0.400pt}}
\put(467.0,253.0){\rule[-0.200pt]{0.400pt}{8.191pt}}
\put(457.0,253.0){\rule[-0.200pt]{4.818pt}{0.400pt}}
\put(457.0,287.0){\rule[-0.200pt]{4.818pt}{0.400pt}}
\put(504.0,248.0){\rule[-0.200pt]{0.400pt}{10.359pt}}
\put(494.0,248.0){\rule[-0.200pt]{4.818pt}{0.400pt}}
\put(494.0,291.0){\rule[-0.200pt]{4.818pt}{0.400pt}}
\put(540.0,174.0){\rule[-0.200pt]{0.400pt}{15.418pt}}
\put(530.0,174.0){\rule[-0.200pt]{4.818pt}{0.400pt}}
\put(530.0,238.0){\rule[-0.200pt]{4.818pt}{0.400pt}}
\put(577.0,172.0){\rule[-0.200pt]{0.400pt}{22.645pt}}
\put(567.0,172.0){\rule[-0.200pt]{4.818pt}{0.400pt}}
\put(567.0,266.0){\rule[-0.200pt]{4.818pt}{0.400pt}}
\put(613.0,222.0){\rule[-0.200pt]{0.400pt}{34.930pt}}
\put(603.0,222.0){\rule[-0.200pt]{4.818pt}{0.400pt}}
\put(603.0,367.0){\rule[-0.200pt]{4.818pt}{0.400pt}}
\put(650.0,159.0){\rule[-0.200pt]{0.400pt}{38.303pt}}
\put(640.0,159.0){\rule[-0.200pt]{4.818pt}{0.400pt}}
\put(640.0,318.0){\rule[-0.200pt]{4.818pt}{0.400pt}}
\put(686.0,251.0){\rule[-0.200pt]{0.400pt}{64.079pt}}
\put(676.0,251.0){\rule[-0.200pt]{4.818pt}{0.400pt}}
\put(676.0,517.0){\rule[-0.200pt]{4.818pt}{0.400pt}}
\end{picture}
}
\put(770,0) {
\setlength{\unitlength}{0.240900pt}
\ifx\plotpoint\undefined\newsavebox{\plotpoint}\fi
\sbox{\plotpoint}{\rule[-0.200pt]{0.400pt}{0.400pt}}%
\begin{picture}(750,540)(0,0)
\font\gnuplot=cmr10 at 10pt
\gnuplot
\sbox{\plotpoint}{\rule[-0.200pt]{0.400pt}{0.400pt}}%
\put(176.0,68.0){\rule[-0.200pt]{0.400pt}{108.164pt}}
\put(176.0,68.0){\rule[-0.200pt]{4.818pt}{0.400pt}}
\put(666.0,68.0){\rule[-0.200pt]{4.818pt}{0.400pt}}
\put(176.0,143.0){\rule[-0.200pt]{4.818pt}{0.400pt}}
\put(666.0,143.0){\rule[-0.200pt]{4.818pt}{0.400pt}}
\put(176.0,218.0){\rule[-0.200pt]{4.818pt}{0.400pt}}
\put(666.0,218.0){\rule[-0.200pt]{4.818pt}{0.400pt}}
\put(176.0,293.0){\rule[-0.200pt]{4.818pt}{0.400pt}}
\put(666.0,293.0){\rule[-0.200pt]{4.818pt}{0.400pt}}
\put(176.0,367.0){\rule[-0.200pt]{4.818pt}{0.400pt}}
\put(666.0,367.0){\rule[-0.200pt]{4.818pt}{0.400pt}}
\put(176.0,442.0){\rule[-0.200pt]{4.818pt}{0.400pt}}
\put(666.0,442.0){\rule[-0.200pt]{4.818pt}{0.400pt}}
\put(176.0,517.0){\rule[-0.200pt]{4.818pt}{0.400pt}}
\put(666.0,517.0){\rule[-0.200pt]{4.818pt}{0.400pt}}
\put(176.0,68.0){\rule[-0.200pt]{0.400pt}{4.818pt}}
\put(176,23){\makebox(0,0){0}}
\put(176.0,497.0){\rule[-0.200pt]{0.400pt}{4.818pt}}
\put(249.0,68.0){\rule[-0.200pt]{0.400pt}{4.818pt}}
\put(249,23){\makebox(0,0){2}}
\put(249.0,497.0){\rule[-0.200pt]{0.400pt}{4.818pt}}
\put(322.0,68.0){\rule[-0.200pt]{0.400pt}{4.818pt}}
\put(322,23){\makebox(0,0){4}}
\put(322.0,497.0){\rule[-0.200pt]{0.400pt}{4.818pt}}
\put(395.0,68.0){\rule[-0.200pt]{0.400pt}{4.818pt}}
\put(395,23){\makebox(0,0){6}}
\put(395.0,497.0){\rule[-0.200pt]{0.400pt}{4.818pt}}
\put(467.0,68.0){\rule[-0.200pt]{0.400pt}{4.818pt}}
\put(467,23){\makebox(0,0){8}}
\put(467.0,497.0){\rule[-0.200pt]{0.400pt}{4.818pt}}
\put(540.0,68.0){\rule[-0.200pt]{0.400pt}{4.818pt}}
\put(540,23){\makebox(0,0){10}}
\put(540.0,497.0){\rule[-0.200pt]{0.400pt}{4.818pt}}
\put(613.0,68.0){\rule[-0.200pt]{0.400pt}{4.818pt}}
\put(613,23){\makebox(0,0){12}}
\put(613.0,497.0){\rule[-0.200pt]{0.400pt}{4.818pt}}
\put(686.0,68.0){\rule[-0.200pt]{0.400pt}{4.818pt}}
\put(686,23){\makebox(0,0){14}}
\put(686.0,497.0){\rule[-0.200pt]{0.400pt}{4.818pt}}
\put(176.0,68.0){\rule[-0.200pt]{122.859pt}{0.400pt}}
\put(686.0,68.0){\rule[-0.200pt]{0.400pt}{108.164pt}}
\put(176.0,517.0){\rule[-0.200pt]{122.859pt}{0.400pt}}
\put(176.0,68.0){\rule[-0.200pt]{0.400pt}{108.164pt}}
\put(556,452){\makebox(0,0)[r]{(1,1)}}
\put(600,452){\raisebox{-.8pt}{\makebox(0,0){$\Diamond$}}}
\put(212,344){\raisebox{-.8pt}{\makebox(0,0){$\Diamond$}}}
\put(249,293){\raisebox{-.8pt}{\makebox(0,0){$\Diamond$}}}
\put(285,272){\raisebox{-.8pt}{\makebox(0,0){$\Diamond$}}}
\put(322,262){\raisebox{-.8pt}{\makebox(0,0){$\Diamond$}}}
\put(358,255){\raisebox{-.8pt}{\makebox(0,0){$\Diamond$}}}
\put(395,241){\raisebox{-.8pt}{\makebox(0,0){$\Diamond$}}}
\put(431,280){\raisebox{-.8pt}{\makebox(0,0){$\Diamond$}}}
\put(467,250){\raisebox{-.8pt}{\makebox(0,0){$\Diamond$}}}
\put(504,241){\raisebox{-.8pt}{\makebox(0,0){$\Diamond$}}}
\put(540,201){\raisebox{-.8pt}{\makebox(0,0){$\Diamond$}}}
\put(577,196){\raisebox{-.8pt}{\makebox(0,0){$\Diamond$}}}
\put(613,382){\raisebox{-.8pt}{\makebox(0,0){$\Diamond$}}}
\put(650,223){\raisebox{-.8pt}{\makebox(0,0){$\Diamond$}}}
\put(212.0,339.0){\rule[-0.200pt]{0.400pt}{2.409pt}}
\put(202.0,339.0){\rule[-0.200pt]{4.818pt}{0.400pt}}
\put(202.0,349.0){\rule[-0.200pt]{4.818pt}{0.400pt}}
\put(249.0,287.0){\rule[-0.200pt]{0.400pt}{2.891pt}}
\put(239.0,287.0){\rule[-0.200pt]{4.818pt}{0.400pt}}
\put(239.0,299.0){\rule[-0.200pt]{4.818pt}{0.400pt}}
\put(285.0,266.0){\rule[-0.200pt]{0.400pt}{2.891pt}}
\put(275.0,266.0){\rule[-0.200pt]{4.818pt}{0.400pt}}
\put(275.0,278.0){\rule[-0.200pt]{4.818pt}{0.400pt}}
\put(322.0,254.0){\rule[-0.200pt]{0.400pt}{3.613pt}}
\put(312.0,254.0){\rule[-0.200pt]{4.818pt}{0.400pt}}
\put(312.0,269.0){\rule[-0.200pt]{4.818pt}{0.400pt}}
\put(358.0,240.0){\rule[-0.200pt]{0.400pt}{6.986pt}}
\put(348.0,240.0){\rule[-0.200pt]{4.818pt}{0.400pt}}
\put(348.0,269.0){\rule[-0.200pt]{4.818pt}{0.400pt}}
\put(395.0,229.0){\rule[-0.200pt]{0.400pt}{5.782pt}}
\put(385.0,229.0){\rule[-0.200pt]{4.818pt}{0.400pt}}
\put(385.0,253.0){\rule[-0.200pt]{4.818pt}{0.400pt}}
\put(431.0,266.0){\rule[-0.200pt]{0.400pt}{6.745pt}}
\put(421.0,266.0){\rule[-0.200pt]{4.818pt}{0.400pt}}
\put(421.0,294.0){\rule[-0.200pt]{4.818pt}{0.400pt}}
\put(467.0,223.0){\rule[-0.200pt]{0.400pt}{13.009pt}}
\put(457.0,223.0){\rule[-0.200pt]{4.818pt}{0.400pt}}
\put(457.0,277.0){\rule[-0.200pt]{4.818pt}{0.400pt}}
\put(504.0,206.0){\rule[-0.200pt]{0.400pt}{17.104pt}}
\put(494.0,206.0){\rule[-0.200pt]{4.818pt}{0.400pt}}
\put(494.0,277.0){\rule[-0.200pt]{4.818pt}{0.400pt}}
\put(540.0,147.0){\rule[-0.200pt]{0.400pt}{26.017pt}}
\put(530.0,147.0){\rule[-0.200pt]{4.818pt}{0.400pt}}
\put(530.0,255.0){\rule[-0.200pt]{4.818pt}{0.400pt}}
\put(577.0,140.0){\rule[-0.200pt]{0.400pt}{26.981pt}}
\put(567.0,140.0){\rule[-0.200pt]{4.818pt}{0.400pt}}
\put(567.0,252.0){\rule[-0.200pt]{4.818pt}{0.400pt}}
\put(613.0,291.0){\rule[-0.200pt]{0.400pt}{44.085pt}}
\put(603.0,291.0){\rule[-0.200pt]{4.818pt}{0.400pt}}
\put(603.0,474.0){\rule[-0.200pt]{4.818pt}{0.400pt}}
\put(650.0,131.0){\rule[-0.200pt]{0.400pt}{44.085pt}}
\put(640.0,131.0){\rule[-0.200pt]{4.818pt}{0.400pt}}
\put(640.0,314.0){\rule[-0.200pt]{4.818pt}{0.400pt}}
\end{picture}
}
\put(950,-25){t}
}
\end{picture}
\caption{$^{1}P_{1}$ Effective masses by (source, sink).}
\label{fig:pplots}
\end{figure}
For the $^{1}S_{0}$ row
fit we choose a value 1.17(5) for the excited state mass
(average of $t_{min}$ = 3,4,6)
and from the matrix fit 1.18(4) (average of $t_{min}$
= 3,4,5). There is then agreement
within errors between the two fits and we choose
1.17(5) as the global average.
For the $^{3}S_{1}$ state there is a significant deterioration in the
Q values over those for the $^{1}S_{0}$
and the fitting errors are slightly larger. This is
presumably a reflection of the additional noise
in the $^{3}S_1$ channel coming from the $^{1}S_0$. For the row
fit a value of 1.19(7) (average for $t_{min}$ = 4,5,6)
is chosen and a value of 1.22(3) (average for $t_{min}$ =3,4,5) from
the matrix fit. A global average for the excited $^{3}S_{1}$ is
chosen to be 1.20(7).
All the fitted values are collected in Table \ref{simresults}.

In Tables \ref{aamplitude} and \ref{bamplitude} are the amplitudes
from the various fits for particular values of $t_{min}/t_{max}$.
The value of $t_{min}/t_{max}$ used was that where the fit for the
first excited state was closest to the average result quoted above.
In both row and matrix fits it was found
that the amplitude for a second excited state ($k$ = 3) is essentially zero.
This indicates that
contamination from higher states in our fits is negligible.
{}From the amplitude results we can see that $n_{sc}$=1 has strongest
overlap with the ground state and $n_{sc}$ = 2 has strongest overlap
with the first excited state, as planned.
Thus our smearing functions are projecting out the required
state and suppressing the others, although our smearing
functions are clearly not optimal. It may be better to use
the output wavefunctions to produce input smearing functions in an
improved calculation.
To illustrate the quality of the multi-exponential fits into
early times we have plotted in Figure 5 effective amplitude plots
with the fitted parameters quoted in Tables \ref{aamplitude} and
\ref{bamplitude}.

\begin{table}
\begin{center}
\begin{tabular}{l|cclllc}
& $N_{exp}$ & $t_{min}/t_{max}$&$\;\;aE_1 \;\;$&$\;\;aE_2\;\;$ &
 $\;\;aE_3\;\;$ & $Q$\\
\cline{1-7}
fits to (1,loc)& 2  & 2/24 & 0.6171(6) &1.172(6) & & $2 \times 10^{-3}$ \\
and (2,loc)    &    & 3/24 & 0.6178(6) &1.16(1) & &0.65  \\
               &    & 4/24 & 0.6176(6) &1.16(1) & &0.64 \\
               &    & 5/24 & 0.6179(7) &1.14(1) & &0.79  \\
               &    & 6/24 & 0.6182(7) &1.21(5) & &0.94   \\
               &    & 7/24 & 0.6183(7) &1.27(8) & & 0.93 \\
               & 3  & 2/24 & 0.6180(7) &1.15(2)  &1.8(6) &0.38  \\
               &    & 3/24 & 0.6177(20) &1.15(4)  &1.8 $\pm$ 15 &0.53  \\
               &    & 4/24 & 0.6181(6) &1.16(2)  &1.8(1.2) &0.79  \\
               &    & 5/24 & 0.6183(7) &1.30(16)  &1.7(6) &0.94  \\
               &    & 6/24 & 0.6183(7) &1.19(8)  &1.8(5) &0.87  \\
               &    & 7/24 & 0.6183(7) &1.25(24)  &1.8(8) &0.85  \\
\cline{1-7}
fits to      & 2  & 3/24 & 0.6185(6) &1.18(2) & & 0.06 \\
(1,1), (1,2) &    & 4/24 & 0.6183(6) &1.17(3) & & 0.15 \\
(2,1), (2,2) &    & 5/24 & 0.6178(6) &1.16(4) & &0.25  \\
             &    & 6/24 & 0.6177(6) &1.08(6) & &0.16  \\
             &    & 7/24 & 0.6181(6) &0.90(6) & &0.42  \\
             & 3  & 3/24 & 0.6180(6) &1.19(2) & 1.6(5) &0.27  \\
             &    & 4/24 & 0.6178(6) &1.14(4) & 2.1(6) &0.23  \\
             &    & 5/24 & 0.6179(6) &1.21(7) & 1.7(6) &0.16  \\
             &    & 6/24 & 0.6180(6) &1.26(11) & 2(1) &0.18  \\
             &    & 7/24 & 0.6181(6) &0.91(6) & 2(1) &0.33  \\
\cline{1-7}
\end{tabular}
\end{center}
\caption{Examples of simultaneous multi-exponential fits to the
$^{1}S_{0}$ using row and matrix fits respectively.}
\label{1s0}
\end{table}

\begin{table}
\begin{center}
\begin{tabular}{l|cclc}
& $N_{exp}$ & $t_{min}/t_{max}$&$\;\;aE_{1} \;\;$& $Q$ \\
\cline{1-5}
fits to (1,loc) & 1 & 5/24 & 0.6188(8) & 0.01 \\
              &   & 6/24 & 0.6184(8) & 0.66 \\
              &   & 7/24 & 0.6183(8) & 0.72 \\
\cline{1-5}
fits to (1,1) & 1 & 4/24 & 0.6184(8) & 0.05 \\
              &   & 5/24 & 0.6181(8) & 0.22 \\
              &   & 6/24 & 0.6181(8) & 0.18 \\
              &   & 7/24 & 0.6182(8) & 0.15 \\
\cline{1-5}
\end{tabular}
\end{center}
\caption{ Examples of single exponential fits to
the $^{1}S_{0}$ .}
\label{1s0single}
\end{table}

\begin{table}
\begin{center}
\begin{tabular}{l|cclllc}
& $N_{exp}$ & $t_{min}/t_{max}$&$\;\;aE_1 \;\;$&$\;\;aE_2\;\;$ &
 $\;\;aE_3\;\;$ & $Q$\\
\cline{1-7}
fits to (1,loc)& 2  & 2/24 & 0.6951(8) &1.247(7) & & $4 \times 10^{-5}$ \\
and (2,loc)    &    & 3/24 & 0.6961(8) &1.23(1) & &0.23  \\
               &    & 4/24 & 0.6958(9) &1.22(2) & &0.23 \\
               &    & 5/24 & 0.6961(9) &1.18(2) & &0.46  \\
               &    & 6/24 & 0.6966(9) &1.21(5) & &0.56   \\
               &    & 7/24 & 0.6968(10)&1.25(8) & & 0.56 \\
               & 3  & 2/24 & 0.6964(9)&1.21(4)  &1.9(9) &0.10  \\
               &    & 3/24 & 0.6957(9) &1.20(4)  &1.9(1.4) &0.17  \\
               &    & 4/24 & 0.6964(10) &1.16(5)  &1.9(1.3) &0.47  \\
               &    & 5/24 & 0.6967(10) &1.22(8)  &1.9(5) &0.55  \\
               &    & 6/24 & 0.6966(7) &1.19(6)  &1.9(3) &0.41  \\
               &    & 7/24 & 0.6969(10) &1.25(16)  &1.9(2) &0.40  \\
\cline{1-7}
fits to      & 2  & 3/24 & 0.6970(8) &1.22(1) & & 0.04 \\
(1,1), (1,2) &    & 4/24 & 0.6967(8) &1.21(3)) & & 0.05 \\
(2,1), (2,2) &    & 5/24 & 0.6965(8) &1.24(5) & &0.07  \\
             &    & 6/24 & 0.6966(8) &1.31(9) & &0.09  \\
             &    & 7/24 & 0.6967(9) &0.95(8) & &0.08  \\
             & 3  & 3/24 & 0.6966(8) &1.23(2) & 1.7(6) &0.08  \\
             &    & 4/24 & 0.6965(8) &1.20(3) & 1.8(6) &0.06  \\
             &    & 5/24 & 0.6964(8) &1.23(4) & 2.0(2.8) &0.04  \\
             &    & 6/24 & 0.6969(8) &1.46(13) & 1.8(1.3)&0.06  \\
             &    & 7/24 & 0.6967(9) &1.00(9) & 1.9(1.3) &0.07  \\
\cline{1-7}
\end{tabular}
\end{center}
\caption{Examples of simultaneous multi-exponential fits to the
$^{3}S_{1}$ using row and matrix fits respectively.}
\label{3s1}
\end{table}

\begin{table}
\begin{center}
\begin{tabular}{l|ccll}
 $\qquad$ Fit $\qquad$& $t_{min}/t_{max}$ & k &$a(n_{sc,sk}=1,k)$ &
 $a(n_{sc,sk}=2,k)$ \\
\cline{1-5}
$N_{exp} = 2$   &4/24 &1& $\;\,$0.681(1) & -0.1188(8)        \\
for $^{1}S_0$   &     &2& $\;\,$0.18(9)  &$\;\,$0.52(2)     \\
\cline{1-5}
$N_{exp} = 2 $  &5/24 &1& $\;\,$0.700(3) & -0.164(1)        \\
for $^{3}S_1$ &     &2& $\;\,$0.29(2)  &$\;\,$0.53(5)     \\
\cline{1-5}
\end{tabular}
\end{center}
\caption{Examples of fit results for amplitudes $a(n_{sc,sk},k)$}
\label{aamplitude}
\end{table}

\begin{table}
\begin{center}
\begin{tabular}{l|ccll}
 $\qquad$ Fit $\qquad$& $t_{min}/t_{max}$ & k &$b(n_{sc}=1,k)$ &
 $b(n_{sc}=2,k)$  \\
\cline{1-5}
 $N_{exp} = 2$      & 4/24 & 1 &0.1037(7) &-0.0184(4)   \\
 for $^{1}S_0$      &        & 2 &0.032(3) &$\;\,$0.064(2)  \\
\cline{1-5}
$N_{exp} = 2$     & 5/24 & 1  &0.103(1) &-0.0253(4)  \\
for $^{3}S_1$     &      & 2  &0.036(7) &$\;\,$0.069(3)  \\
\cline{1-5}
\end{tabular}
\end{center}
\caption{Examples of fit results for amplitudes $b(n_{sc},k)$}
\label{bamplitude}
\end{table}

For the P and D states multiple exponential fits
are not possible because we have included only
the ground state smearing function in the simulation.
Instead a single exponential fit was performed to the
$(n_{sc},n_{sk})$ = $(1,1)$ meson propagators of the
$^{1}P_{1}$ and $^{1}D_{2}$. The results are shown in Tables
\ref{1p1} and \ref{1d2}. Reasonable errors are obtained at
$t_{min}$ values of 6 where single exponential fits were
acceptable for the $S$ states.
Ratio fits were also done to the $^{1}S_{0}$ in both
cases but the results and errors remained the same
showing there is no correlation between these
states and the $^{1}S_{0}$.
To isolate the ground state early on and
achieve better errors higher
radial smearing functions need to be added. Work
has begun on this for the $^{1}P_{1}$ state.

\begin{table}
\begin{center}
\begin{tabular}{l|cclc}
& $N_{exp}$ & $t_{min}/t_{max}$&$\;\; aE_{1} \;\;$& $Q$ \\
\cline{1-5}
fits to (1,1) & 1& 3/24 & 1.059(4) & 0.45 \\
              &  & 4/24 & 1.052(5) & 0.68 \\
              &  & 5/24 & 1.049(7) & 0.66 \\
              &  & 6/24 & 1.046(9) & 0.62 \\
              &  & 7/24 & 1.048(14) &0.55 \\
\cline{1-5}
\end{tabular}
\end{center}
\caption{Example of a $^{1}P_{1}$ fit.}
\label{1p1}
\end{table}

\begin{table}
\begin{center}
\begin{tabular}{l|cclc}
& $N_{exp}$ & $t_{min}/t_{max}$&$\;\; aE_{1} \;\;$& $Q$ \\
\cline{1-5}
fits to (1,1) & 1& 3/24 & 1.35(1) & 0.62 \\
              &  & 4/24 & 1.32(2) & 0.77 \\
              &  & 5/24 & 1.30(3) & 0.78 \\
              &  & 6/24 & 1.26(5) & 0.78 \\
              &  & 7/24 & 1.26(9) & 0.72 \\
\cline{1-5}
\end{tabular}
\end{center}
\caption{Example of a $^{1}D_{2}$ fit. }
\label{1d2}
\end{table}

\begin{figure}
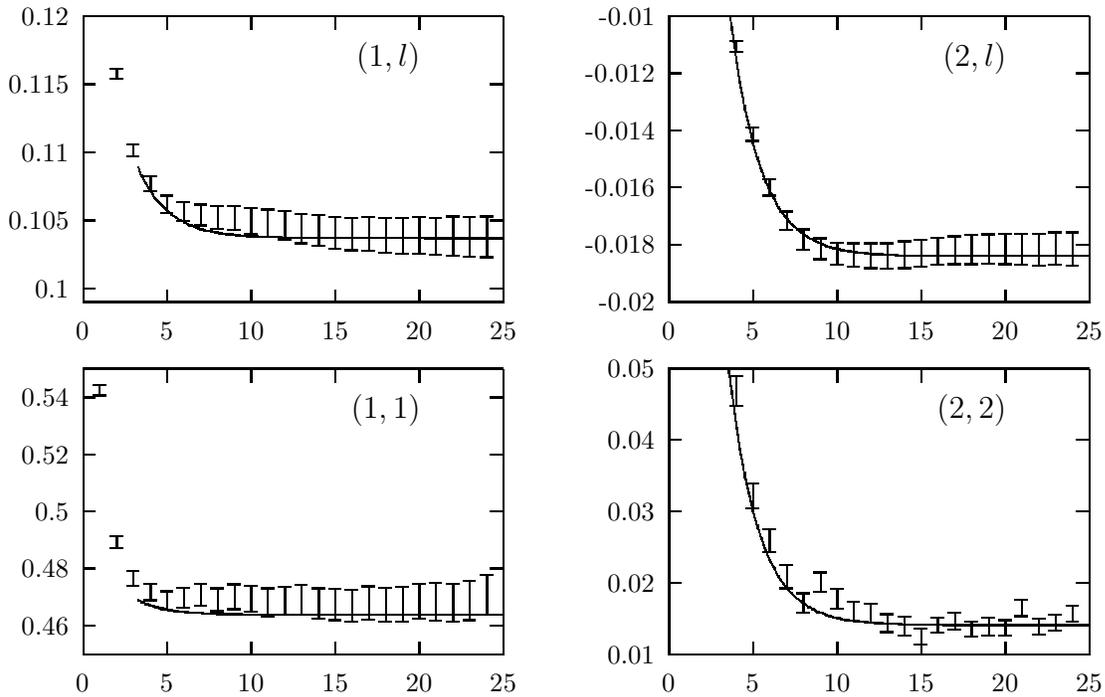

%
}
}
\end{picture}
\caption{$^{1}S_{0}$ Effective amplitudes
$G(t)\; \cdot e^{E_1 \cdot t} $
from two-exponential row fits (1,$\ell$) (2,$\ell$)
and two-exponential matrix fits (1,1) (2,2)
with $t_{min} = 4 , t_{max} = 24$ .}
\end{figure}

\subsection{Fits to Spin Splittings}

As described earlier, spin splittings are very dependent on
the tadpole improved coupling
constants $c_{i}$. This makes the spin-splittings a good test of
the tadpole-improvement scheme. It is also true that potential models
find it hard to produce spin-splittings in agreement with experiment
so we would hope that they are also a good test of the differences
between a full calculation in QCD, such as ours, and a potential model.

Since meson correlation functions of given $l$ from
 the same configuration are highly
correlated we produce a bootstrap ensemble of ratios of correlation functions
to find spin splittings. {}From this we fit to a single exponential
\begin{equation}
Ratio(t) = A e^{-\delta E t}
\label{ratio}
\end{equation}
We use correlation functions with $(n_{sc}, n_{sk})$ = (1,1) and bin
on time and spatial origin. We
find very high Q values in general. Shown in Table \ref{ratiofits}
are values obtained for various combinations of spin-splittings using
equation (\ref{ratio}).
The $\delta E$ obtained for the $^{3}S_1$ to $^{1}S_0$ ratio fit is in
agreement with that obtained from the separate row and matrix fits of
Tables \ref{1s0} and \ref{3s1}.
To estimate $\delta E$ for higher radial excitations
we have used a correlated $\delta E$ fit. This is a
fit to the form

\begin{eqnarray}
G_{meson\;A}(n_{sc},loc;t)  &=&  \sum_{k=1}^{N_{exp}}\; c_A(n_{sc},k)\,
e^{-E_k^A \cdot t } \nonumber \\
G_{meson\;B}(n_{sc},loc;t) &=& c_B(n_{sc},1)\,e^{-(E_1^A+ \delta E) \cdot t}
\; + \;\sum_{k=2}^{N_{exp}}\; c_B(n_{sc},k)\,e^{-E_k^B \cdot t }
\end{eqnarray}
with $n_{sc} = 1,2$ for each meson. The results shown in
Table \ref{corrdeltae} show that the $^{3}S_{1} - ^{1}S_{0}$
splitting can be obtained at early
times with smaller errors than in the ratio
fit. Presumably extra excited states have been absorbed in the extra terms
in the correlated fit.
We are unable to obtain a clear signal for a 2S hyperfine splitting
although the correlated $\delta E$ fit above and the individual
matrix fits give an indication of such a splitting at early times.

\begin{table}
\begin{center}
\begin{tabular}{l|ccccc}
Splitting & $N_{exp}$ & $t_{min}/t_{max}$ & $a \delta E$ &  $Q$\\
\cline{1-5}
$^{3}S_{1}\;-\;^{1}S_{0}$ & 1 & 4/24  &  0.0794(3)   & $4.0\times 10^{-5}$\\
                     &   & 6/24  &  0.0784(4)   & 0.35 \\
                     &   & 8/24  &  0.0784(4)   & 0.32  \\
                     &   & 10/24 &  0.0783(5)   & 0.21 \\
                     &   & 12/24 &  0.0778(6)   & 0.25 \\
\cline{1-5}
$^{3}P_{2E}\;-\;^{3}P_{0}$   & 1 &  3/13  &  0.090(2)   & 0.94  \\
                        &  &  4/13  &  0.089(4)   & 0.91  \\
                        &  &  5/13  &  0.090(6)   & 0.85  \\
                        &  &  6/13  &  0.086(9)   & 0.81  \\
\cline{1-5}
$^{3}P_{2E}\;-\;^{3}P_{1}$ & 1 & 3/13 & 0.045(1) & 0.99 \\
                       &  & 4/13 & 0.046(3) & 0.99 \\
                       &  & 5/13 & 0.045(4) & 0.99 \\
                       &  & 6/13 & 0.044(6) & 0.97 \\
\end{tabular}
\end{center}
\caption{Examples of ratio fits for spin-splittings }
\label{ratiofits}
\end{table}

\begin{table}
\begin{center}
\begin{tabular}{ccccccc}
& $t_{min}/t_{max}$ & $1^{1}S_{0}$ & $2^{1}S_{0}$ & $1^{3}S_{1}-
1^{1}S_{0}$ & $2^{3}S_{1}$ &  $Q$ \\
\cline{1-7}
& 3/24 & 0.6179(6) &  1.17(1) & 0.0779(3) & 1.23(1) & 0.29 \\
& 4/24 & 0.6178(6) &  1.17(1) & 0.0778(3) & 1.24(2) & 0.33 \\
& 5/24 & 0.6180(6) &  1.16(2) & 0.0777(3) & 1.19(2) & 0.72 \\
& 6/24 & 0.6183(6) &  1.20(4) & 0.0780(4) & 1.20(4) & 0.88 \\
& 7/24 & 0.6183(7) &  1.20(6) & 0.0781(4) & 1.20(6) & 0.82 \\
& 8/24 & 0.6184(7) &  1.16(11)& 0.0781(4) & 1.25(13)& 0.76  \\
\cline{1-7}
\end{tabular}
\end{center}
\caption{Example of correlated $\delta E$ fit for the
$^{3}S_{1}$ and $^{1}S_{0}$ states }
\label{corrdeltae}
\end{table}

\begin{table}
\begin{center}
\begin{tabular}{c|c}
& Simulation Results \\
\cline{1-2}
$1^{1}S_{0}$ & 0.6182(7) \\
$1^{3}S_{1}$ & 0.697(1) \\
$2^{1}S_{0}$ & 1.17(5) \\
$2^{3}S_{1}$ & 1.20(7) \\
$1^{1}P_{1}$ & 1.05(1) \\
$1^{1}D_{2}$ & 1.30(4) \\
\cline{1-2}
$^{3}S_{1} - ^{1}S_{0}$ & 0.0782(4) \\
$^{3}P_{2} - ^{3}P_{0}$ & 0.088(8) \\
$^{3}P_{2} - ^{3}P_{1}$ & 0.044(5) \\
$^{3}P_{1} - ^{3}P_{0}$ & 0.044(3) \\
$^{3}P_{CM} - ^{1}P_{1}$ & 0.010(1) \\
\cline{1-2}
\end{tabular}
\end{center}
\caption{Fitted dimensionless energies.}
\label{simresults}
\end{table}

\section{Comparison with Experiment}

In Table \ref{simresults} we give the dimensionless splittings
obtained from our fitting procedure.
To compare simulation results to experiment it is necessary
to fix the scale $a^{-1}$. We choose the spin-averaged 1P-1S splitting to
do this. By spin-averaged splitting we mean the splitting
between spin-averaged states. The spin-averaged S state has mass
$0.25 \times [3m(^3S_1)+m(^1S_0)]$. The spin-averaged P state has either
the mass of the $^1P_1$ or mass $ 1/9 \times [5m(^3P_2)+3m(^3P_1)+m(^3P_0)]$.
These two P masses are the same in potential models
and experimentally
they do seem to be very close although the mass of the $^{1}P_1$ needs
confirmation \cite{PDG}.
In our simulation the two masses are slightly different (see
Table \ref{simresults}). We will
use $m(^1P_1)$ because of the previously noted disagreement with
experiment in the P fine structure.
The difference in value of the $a^{-1}$s obtained is within the
statistical error.

 The spin-averaged 1P-1S splitting has the advantage of being
independent of any errors in spin-dependent terms
and of being experimentally known to be
independent of the heavy quark mass in the $c$, $b$, region.
This gives much less systematic uncertainty than, for example,
in light hadron spectrum determinations of $a^{-1}$.
In the $\Upsilon$ spectrum calculation \cite{spectrum}
it was possible to see a difference in $a^{-1}$ between that
fixed from the 2S-1S splitting and that fixed from the 1P-1S splitting.
Here both our statistical error on the 2S state and our expected
systematic error from relativistic corrections are too large for
this to be possible.

Using the values in Table \ref{simresults}
we find
$a^{-1}$ = 1.23(4) GeV from the 1P($^{1}P_1$)-1S splitting.
In Table \ref{results} we compare the splittings obtained
from this simulation with experimental results. The results are plotted
in Figures (\ref{fig:cspinav}) and (\ref{fig:cspinp}).
 It is important to remember that there is
a potential $30-40$~MeV systematic error in all splittings coming
from relativistic corrections not included in the heavy quark action.
Table \ref{results} and the figures do not include the statistical
error in $a^{-1}$ in their quoted errors since all the splittings
are correlated. Table \ref{results} does, however, include
this error for the hyperfine splitting since this is very
sensitive to shifts in the bare quark mass allowed by uncertainties
in $a^{-1}$ (the hyperfine splitting behaves as $1/M_Q$ in perturbation
theory, see equation (16) below). Using the $\chi_c$ average for 1P would give
$a^{-1}$ = 1.20, at the lower end of the range for $a^{-1}$ from
the $^{1}P_1$.

\begin{table}
\begin{center}
\begin{tabular}{c|ll}
 & Simulation Results [GeV]  &  Experiment [GeV]  \\
\cline{1-3}
$2{^1S}_{0}\; - \;1{^1S}_{0}$  & 0.68(6) &   \\
$2{^3S}_{1}\;-\;1{^3S}_{1}$  &  0.62(8) & 0.589(1) \\
${^1D}_{2}\; - \;{^1S}_{0}$ & 0.84(5)  \\
${^3D}_{1}\; - \;{^1S}_{0}$ &         & 0.791(3) \\
\cline{1-3}
${^3S}_1\;-\;{^1S}_0$ &0.096(2)  & 0.118(2) \\
${^3P}_2\;-\;{^3P}_0$ &0.11(1)  & 0.141(1)  \\
${^3P}_2\;-\;{^3P}_1$ &0.054(6)   & 0.0456(1)  \\
${^3P}_{CM}-{^1P}_1$  &0.012(1)   & $0.0008(3)^{*}$  \\
\cline{1-3}
\end{tabular}
\end{center}
\caption{NRQCD spectrum results
and comparison with experiment for $a^{-1}$ = 1.23 GeV and
$a\mbare$ = 0.8. $^{*}$ requires confirmation.}
\label{results}
\end{table}

As discussed earlier, the statistical error on the 2S state is too
large to see any significance in the fact that it is  slightly
higher than experiment.
 The
direction of the slight disagreement
 is the same as that for the $\Upsilon$
spectrum \cite{spectrum}. There it seems clear that
the correction of ${\cal O} (a^{2})$ errors in the gluon action and
unquenching will produce agreement with experiment \cite{alphas,bielefeld}.
To test this for the $\Psi$ we will need to reduce the statistical errors
and systematic errors from the heavy quark action in the 2S state.

The expected shift in the 1S state from gluonic $\cal{O}$$(a^2)$
effects is 0.006 in lattice units for this simulation. This is
calculated either perturbatively from the wavefunction at the
origin \cite{alphas} or non-perturbatively using a lattice potential
model \cite{andrew}. It is less than the shift for the $\Upsilon$
at the same value of $\beta$ since the $J/\Psi$ is larger.
The 1P state does not shift, since it is
not sensitive to perturbations at the origin. The change in the 1P-1S
splitting would then cause the derived $a^{-1}$ to change
upwards to 1.25(4) GeV if
gluonic $\cal{O}$$(a^2)$ effects were corrected. This is still within
$1 \sigma$ of the original value. The expected shift in the $\psi'$
state is 0.005 so the change in the 2S-1S splitting would be completely
negligible compared to its statistical error.

The value for $a^{-1}$ is clearly different from that from
$\Upsilon$\cite{bielefeld} or light hadron \cite{weingarten}
 spectroscopy at the same value of $\beta$. In the quenched approximation
we would expect $a^{-1}_{b\overline{b}} > a^{-1}_{c\overline{c}} >
a^{-1}_{m_{\rho}}$, reflecting the ordering of the
momentum scales appropriate to the different quantities. In current
results,
the first inequality holds but the second one does not\cite{bielefeld};
this may reflect $\cal{O}$$(a)$ errors in present light hadron
spectroscopy. Further calculations at different values of $a$
will resolve this problem.

The $^{1}D_2$ state whose mass we have calculated
is rather higher than that found for the $\psi(3770)$, thought to
be a $^{3}D_1$ state. {}From the spin splittings alone you would expect
this difference. The $\psi(3770)$ is also above threshold for decay
to $\overline{D}D$ so quenching might have a significant effect on
masses in this region, although
the ratio of the width of the $\psi(3770)$ to
 its mass is still less than 1\%.
The $^{3}D_{1}$ has the same $J^{PC}$ quantum numbers as the $^{3}S_{1}$
and will appear as a third excited state in that channel.
In order to observe such a state the cross-correlation between the
meson correlators $^3{S}_{1}$ and the $^3{D}_{1}$ would have to be calculated
and we have not attempted to do this here.

Values for the wave function at the origin can be obtained
as discussed in ref. \cite{spectrum}. If we include the $(loc,loc)$
correlation function in a multi-exponential row fit we obtain
a value of $a^{3/2}\psi(0)$ for the $J/\Psi$ of 0.1535.
This method does not yield a stable value for the excited states
since the $(loc,loc)$ correlation function does not
distinguish different states very readily.
A better method is take a ratio of amplitudes from row and matrix
fits \cite{spectrum}. We use $b(n_{sc},m)/a(n_{sc},m)$ and
concentrate on the diagonal entries i.e. $n_{sc}$ = $m$ =1 for
$J/\Psi$ and $n_{sc}$ = 2 for $\psi'$. This gives
$a^{3/2}\psi(0)$ = 0.148(2) for $J/\Psi$ and 0.13(1) for $\psi'$.

The leptonic width can be calculated from $\psi(0)$ using the
Van-Royen Weisskopf formula\cite{leptonic} at leading order. We obtain
5.4(5) keV for the $J/\Psi$ in good agreement with the
experimental value of 5.3(3) keV. The error we quote is
dominated by the error in $a^{-1}$ since this appears cubed.
In principle we expect large
corrections ($\approx$ 30\%) to our value when a current correctly matched to
the continuum current is included, instead of the na\"{\i}ve lowest
order current that we have used. We should apply small-components
corrections to the current \cite{collins} as well as a lattice-to-continuum
renormalisation. The agreement with experiment should
thus not be taken to be very significant at this stage.
For the $\psi'$ the agreement with experiment is not so good. The
experimental value is less than half of that for the 1S and yet
we obtain a ratio of 0.7 to the 1S. This trend for excited states
to have too large a value for $\psi(0)$  is again similar to that
found in the $\Upsilon$ case. On improving the systematic error
in our currents we would hope to notice an improvement here unless it is
a feature of the quenched approximation.

Spin splittings have been calculated for the
ground S and P states.
These are shown in detail in Figure (\ref{fig:cspinp}). The
agreement with experiment is good within expected systematic
errors of $30 - 40$ MeV.
 This would not be possible without
tadpole-improvement of the spin-dependent terms. It was clear
from the $\Upsilon$ spectrum\cite{spectrum} that splittings without tadpole
improvement were about half the size of those with tadpole
improvement. This would be an even bigger effect here where
$\beta$ and $u_0$ are smaller.
There is nevertheless some disagreement with experiment in
Figure (\ref{fig:cspinp}), and it
is useful to find the source of this. There are sufficient
experimental results for charmonium that the system provides a
good test of the systematic removal of sources of error.

{}From Table \ref{results}
we can see that
the hyperfine splitting $M(^3S_1) - M(^1S_0)$ has a very small
statistical error. The difference
from experiment then shows up clearly and is presumably a
result of our systematic errors. There is again a $30-40$~MeV systematic error
from higher order relativistic, discretisation and radiative corrections
to the heavy quark action. This would be quite sufficient to explain
the difference. Relativistic corrections are documented in \cite{nrqcd}.
 The radiative corrections are
${\cal O} (g^{2})$ corrections to the coefficient of the
$\sigma \cdot \vec{B}$ term beyond tadpole improvement.
 The discretisation errors are ${\cal O} (a^{2})$
errors in the $\vec{B}$ field and the hyperfine splitting
is rather sensitive to these, as discussed below.
We also expect quenching to have a significant effect, however.
A comparison of $\Upsilon$ results on quenched and unquenched configurations
shows an increase in the hyperfine splitting when unquenched (to 3 flavours)
of between 30\% and 50\% \cite{alphas,bielefeld}. This can be explained largely
on the basis of the
difference between quenched and unquenched coupling constants and
wavefunctions
appearing in the perturbative formula for the hyperfine splitting,
\begin{equation}
\Delta M_{\rm hfs} =
\frac{32\pi\,\alpha_s(M_Q)}{9\,M_c^2}\,\left|\psi(0)\right|^2 .
\end{equation}
For the $J/\Psi$ case we might expect a similar shift
of the hyperfine splitting on going to the full theory and this
again would be sufficient to explain fully the deviation from
experiment. One problem here is that the perturbative formulae
are not quite as reliable as in the $\Upsilon$ case \cite{alphas}.

A calculation of the $\overline{c}c$ hyperfine splitting by the
Fermilab group \cite{junko} gives a somewhat smaller value than ours.
They use an improved Wilson fermion action for the heavy quarks and this
approach has different systematic errors than ours.

The case of the P state fine splittings is much more complicated, with
an expected interplay of short and long range effects.
In a potential model approach \cite{peskin}
two terms contribute - one proportional to
 $\langle \vec{L}\cdot\vec{S} \rangle$ and
the other proportional to $\langle S_{12} \rangle$ where
$S_{12} = 4[3
(\vec{s_1}\cdot\hat{n})(\vec{s_2}\cdot\hat{n}-\vec{s_1}\cdot\vec{s_2})]$. Here
$s_1$ and $s_2$ are the spins of the heavy quarks and $\hat{n}$ is
an arbitrary unit vector. Evaluating these expectation values
for $^3P_0$, $^3P_1$ and $^3P_2$ states of equal mass quarks allows us to
compare
ratios of the splittings, since the
the expectation values of potentials that accompany these
terms are the same for all $P$ states.
A useful ratio \cite{peskin} is
\begin{equation}
r = \frac {M(\chi_2) - M(\chi_1)} {M(\chi_1)-M(\chi_0)}
\end{equation}
Experimental values are
0.48(1) for $c\overline{c}$, 0.66(2) for $b\overline{b}$ (1P) and
0.58(3) for $b\overline{b}$ (2P).
{}From a comparison of possible potentials to experiment the conventional
picture emerges in which
the spin-orbit potential appearing with $\vec{L}\cdot \vec{S}$
has both short and long range pieces, whereas the tensor potential
appearing with $S_{12}$ has only a short range piece.
The short-range pieces can be related to 1-gluon exchange in perturbation
theory and behave like $1/R^3$. The long-range piece comes from
the scalar confining potential.
Spin-dependent potentials can be extracted on the lattice
from expectation values of Wilson loops with $E$ and $B$
field insertions along the time lines on either side.
 There it becomes clear that
the `same-side' spin-orbit potential is long-range, whereas
the `opposite-side' is short-range, as is the tensor potential\cite{michael1}.

We can extract values for the above ratio $r$ of P spin splittings
from our simulation and we find
1.2(2) for $c\overline{c}$, clearly too large.
 The $b\overline{b}$ (1P) result at $\beta$ = 6.0 \cite{spectrum} is
0.7(3), which is consistent with experiment, but at $\beta$ = 5.7 we
obtain 1.4(4) \cite{us_in_progress}.
It seems likely then that the disagreement with experiment arises
from discretisation errors.  At low $\beta$ the predominant spin-dependent
potential is the long-range spin-orbit piece, the shorter range
pieces are not well resolved (compare \cite{michael2} and \cite{bali},
for example). A pure $\vec{L}\cdot\vec{S}$ potential would give a
value for $r$ of 2 \cite{peskin} (the pure tensor would give $-0.4$).
In potential model language the long-range $\vec{L}\cdot\vec{S}$ term has
undue dominance in our simulation.
We also find that the overall size of the P spin splittings,
set by $M(\chi_2)-M(\chi_0)$, is too small.
 For $b\overline{b}$ at
$\beta$ = 6.0 this splitting was on the low side but in
agreement with experiment within the error \cite{spectrum}.
For $b\overline{b}$ at $\beta$ = 5.7 we obtain a result which is much too
small \cite{us_in_progress}.
Future calculations will
concentrate on correcting discretisation errors to see if the results
for charmonium at low $\beta$ improve.

Another possible discretisation error shows up in the fact that the
centre of mass of the $^3P$ states comes out above
the $^1P_1$. This happens both for this calculation and
that of the $\Upsilon$ spectrum \cite{spectrum}, but
in both cases at a level within the expected systematic errors.
One might expect, for example, that the hyperfine $\vec{S}\cdot\vec{S}$
interaction would contribute such a term to $P$ states even in the absence of
a wavefunction at the origin (see equation (16)) if the
$B$ field was smeared out over a plaquette as it is here.
Experimental evidence so far indicates that there
is no such splitting \cite{PDG}, although it awaits
confirmation.

It seems likely that errors from the quenched approximation (and
from discretisation) are not so
large for the P fine structure as for the S hyperfine splitting
because the latter is determined by very short
range phenomena.
The hyperfine splitting can be thought
of as resulting from delta-function $\vec{S_1}\cdot \vec{S_2}$ potential
at the origin (see equation (16)), where S states have
significant wave function.
The quenched approximation causes larger effects at short
distance scales because it appears, perturbatively,
as an incorrect running of the coupling constant $\alpha(R)$ down
to the origin from some $\overline{R}$ which is the important separation
for quark and antiquark
in the 1P-1S splitting which is used to set the scale.
P states have no wavefunction at the origin
and in addition the short-range pieces of
the relevant spin-dependent potentials have longer range than the delta
function hyperfine potential.
 This should mean that
the P fine structure can be determined accurately in a quenched
calculation by a systematic improvement on this
calculation, without having to
unquench.

\section{Conclusions}

This represents a first calculation of the $c\overline{c}$ spectrum
using NRQCD with spin-dependent terms.
 We include the leading relativistic and discretisation
errors with tadpole-improved coefficients.
We find a value of the lattice spacing from the 1P-1S splitting which
is different from that of the $\Upsilon$ on the same
configurations\cite{us_in_progress}. This is a clear indication of an effect
from
the quenched approximation. Another effect seen in the $\Upsilon$ spectrum
itself, the difference in $a^{-1}$ from the 2S-1S and 1P-1S splittings,
is not visible here above the statistical noise in the 2S state.

With tadpole-improvement, the spin splittings
agree with experiment at the level of the
systematic error that we expect. The trend of these systematic errors
is the same as that for the $\Upsilon$ spectrum and we would
expect that, on including higher order terms, we could obtain
better agreement. It seems likely that the major
errors at present are discretisation effects and future calculations
will correct for these.
 One very good feature of the $c\overline{c}$ spectrum
is that all the radial ground state S and P masses are
 known experimentally and so they
can be used to gauge the effect of systematic improvement.
Further calculations of the $c\overline{c}$ spectrum on lattices
of different lattice spacing and on unquenched configurations would
also provide useful checks of the systematic errors. A value for
$\alpha_s$ could be extracted from the 1P-1S splitting in the
same way that it was done using the $\Upsilon$ calculation \cite{alphas}
and a comparison with results from Wilson fermions \cite{fnal,aoki} made.

Calculations of the $B_c$ spectrum combining $b$ and $c$ propagators
on these configurations will be reported shortly \cite{us_in_progress}.

\vspace{4mm}

{\bf Acknowledgements}
This calculation was performed at the Atlas Centre under grant GR/J18927 from
the UK PPARC. AJL is also grateful to PPARC for a studentship,
CTHD for support under grant GR/J21231 and JSl for a Visiting Fellowship to
Glasgow
while this work was being completed.
This work was supported in part by grants from the U.S.Department of
Energy (DE-FC05-85ER250000, DE-FG05-92ER40742, DE-FG02-91ER40690), and
the National Science Foundation.
We thank UKQCD for
making their configurations available to us, and in particular David
Henty who helped us to read them.  Finally, we wish to
acknowledge fruitful discussions with Aida El-Khadra, Paul Mackenzie,
 Colin Morningstar and Beth Thacker.

\end{document}